\documentclass[final,1p,times]{elsarticle} 

\usepackage{graphicx}
\usepackage{amssymb} 
\usepackage{amsthm} 
\usepackage{lineno}

\newcommand{\be}{\begin{equation}}
\newcommand{\ee}{\end{equation}}
\newcommand{\bea}{\begin{eqnarray}}
\newcommand{\eea}{\end{eqnarray}}
\newcommand{\bean}{\begin{eqnarray*}}
\newcommand{\eean}{\end{eqnarray*}}
\newcommand{\vecp}{{\mathbf p}}
\newcommand{\md}{\mathrm{d}}
\newcommand{\vecx}{{\mathbf x}}
\newcommand{\vecnull}{{\mathbf 0}}

\newcommand{\Jpsi}{J/\psi}

\journal{Nuclear Physics A} 

\begin{document}

\begin{frontmatter} 

\title{Momentum dependences of charmonium properties from lattice QCD}

\author{Heng-Tong Ding}

\address{Physics Department, Brookhaven National Laboratory, Upton, NY 11973, USA}

\begin{abstract}

Charmonia produced in initial hard parton scatterings during heavy ion collisions move with respect to the
medium rather than flow with the medium. Lattice studies suggest that charmonium bound states at the rest are dissociated at $T\gtrsim 1.5~T_c$. 
We present results on momentum dependences of charmonium properties in a hot medium from lattice QCD Monte Carlo simulations. 
The dispersion relation of the screening mass and the change of correlation and spectral functions at various temperatures and momenta are
discussed.

\end{abstract} 

\end{frontmatter} 


\section{Introduction}

In medium hadron properties can provide useful information about the hot medium formed during 
heavy ion collisions~\cite{Ding:2012ar}, in particular heavy quarkonium can probe
the deconfinement aspect of QGP and serve as a QGP thermometer~\cite{Matsui:1986dk}. In recent years there have been great progress 
in understanding the properties of heavy quarkonium at rest in the medium from lattice QCD Monte Carlo simulations
and heavy quark effective theories~\cite{Kaczmarek:2012ne,Rothkopf:2012et}. However, quarkonia that are produced from initial hard parton
collisions are moving with respect to medium since $\Jpsi$ almost does not flow with the medium up to $p_\perp \sim 8 $ GeV/c ~\cite{Tang:2011kr,Massacrier:2012yj}.
Properties of moving heavy quarkonia have been investigated by modeling $\Jpsi$ traveling in a finite temperature gas of mesons ~\cite{Haglin:2000ar} and also in the weak coupling regime using effective theories~\cite{Escobedo:2011ie,Aarts2012} as well as in the strong coupling regime using the AdS/CFT analogy~\cite{Liu:2006nn}. These studies all suggest that the bound state peaks get attenuated~\cite{Haglin:2000ar,Escobedo:2011ie,Liu:2006nn} rather than unmodified~\cite{Chu:1987sy} at large momenta.

The moving $\Jpsi$ with respect to a thermal heat bath can be treated as an external particle in an equilibrium background
and its properties are thus suitable to be investigated on the lattice.  From lattice QCD Monte Carlo simulations one can calculate
various correlation functions, i.e. spatial correlation functions $G_H(z,\vecp_{\perp},\omega_n) $ and temporal correlation functions $G_H(\tau,\vecp)$.
They are 
related with the spectral function $\rho_H(\omega,\vecp,T)$ in the following
{\small
\bea
G_H(\tau,\vecp) &=& \sum_{x,y,z} \exp(-i \vecp \cdot \vecx)\left \langle J_H(0,\vecnull)J_H^{\dagger}(\tau,\vecx)\right \rangle = \int_0^{\infty}\frac{\md\omega}{2\pi}\, \,\,\frac{\cosh(\omega(\tau-1/2T))}{\sinh(\omega/2T)}\,\, \rho_H(\omega,\vecp,T), \\
G_H(z,\vecp_{\perp},\omega_n) &=&  \sum_{x,y,\tau} \exp(-i \tilde{\vecp} \cdot \tilde{\vecx})\left \langle J_H(0,\vecnull)J_H^{\dagger}(\tau,\vecx) \right\rangle =\int_{-\infty}^{\infty} \frac{\mathrm{d}p_z}{2\pi}\, \exp (ip_z\,z) \int_{0}^{\infty} \frac{\mathrm{d}\omega}{\pi}\frac{\omega\, \rho_H(\omega,\vecp, T)}{\omega^2+\omega_n^2}\, ,
\label{cor_spf_relation}
\eea
}
where $J_H$ is a local meson operator and has a form of $\bar{q}\Gamma_H q$. $\Gamma_H = \gamma_5$ stands for the pseudo scalar channel ($\eta_c$) and 
$\Gamma_H = \gamma_\mu$ stands for the vector channel ($\Jpsi$) . $\tau$ is the Euclidean time and restricted by the inverse temperature, i.e. $\tau T\leq 1$, $z$ is the spatial distance and is not limited
by the temperature, $\vecp_\perp=(p_x,p_y)$ is the transverse momentum, $\tilde{\vecp}=(\vecp_\perp,\omega_n)$, $\omega_n=2\pi n T$ are the Matsubara frequencies and $\tilde{\vecx}=(x,y,\tau)$.

All information on quarkonium properties are enclosed in the spectral function and thus the task in principle is straightforward, i.e.  to extract the spectral function
from temporal or spatial correlation functions. Normally one extracts spectral functions from the temporal correlation functions as obviously the spatial correlation 
function has a more intricate relation with the spectral function than the temporal one.

\section{Lattice QCD results}

The charmonium properties at vanishing momentum has been recently investigated on large and fine quenched lattices in Ref.~\cite{Ding:2012sp}, in which both P wave states ($\eta_c$ and $\Jpsi$)
and S wave states ($\chi_{c0}$ and $\chi_{c1}$) are found to dissolve at $T\gtrsim 1.5~T_c$. In this proceeding we report results on the momentum dependences of charmonium properties
measured on the finest lattice used in Ref.~\cite{Ding:2012sp}.

At nonzero ``momentum" ($\vecp_{\bot}\neq0$ or $\omega_n \neq0$), the exponential decay at large distance of the spatial correlator may be described by an energy $E_{sc}$
\be
G(z,\vecp_\bot,\omega_n)\sim \exp(-E_{sc}z),~~~~E^2_{sc}=\vecp_{\bot}^2+\frac{\omega_n^2}{A^2} + m_{sc}^2,
\label{def:scr_mass_dispersion_relation}
\ee 
where $m_{sc}$ is the screening mass which can differ from the pole mass if $A(T)\neq 1$. It is worth noting that the above ansatz is based on the dispersion relation in the continuum limit.
 \begin{figure}[htpl]
      \begin{center}
    \includegraphics[width=.33\textwidth]{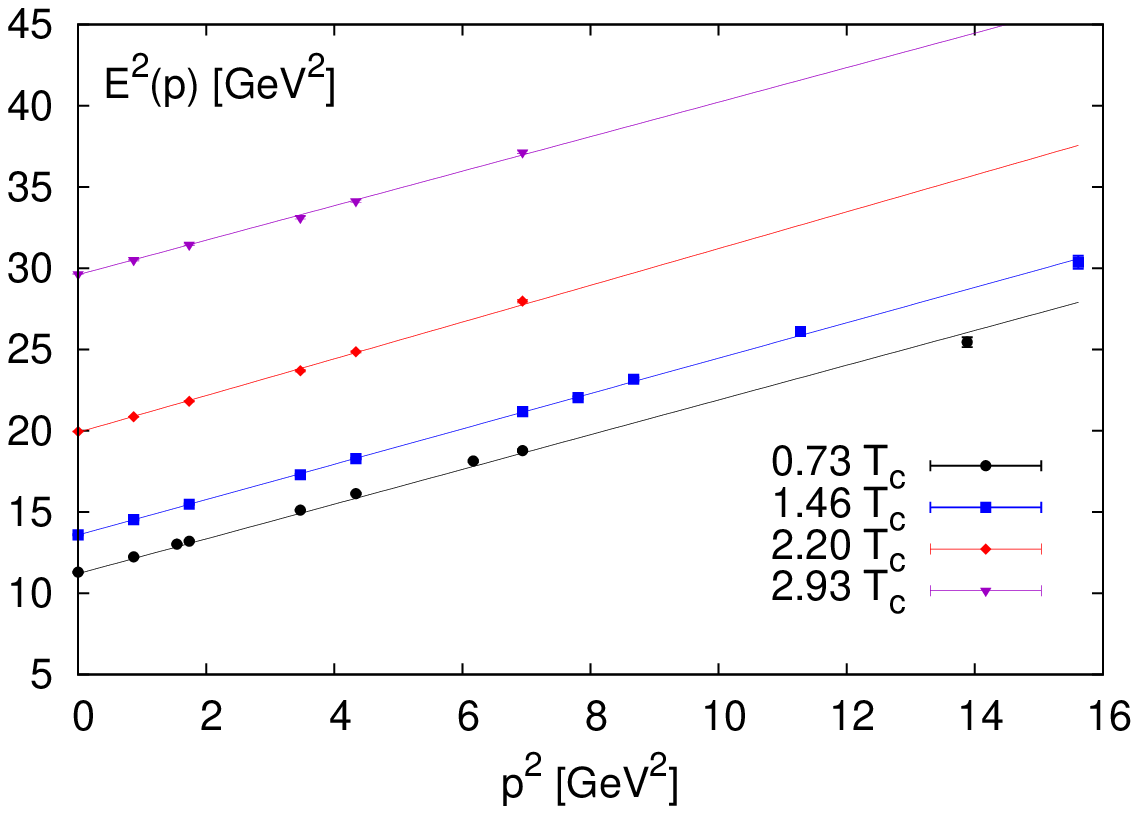}~ \includegraphics[width=.33\textwidth]{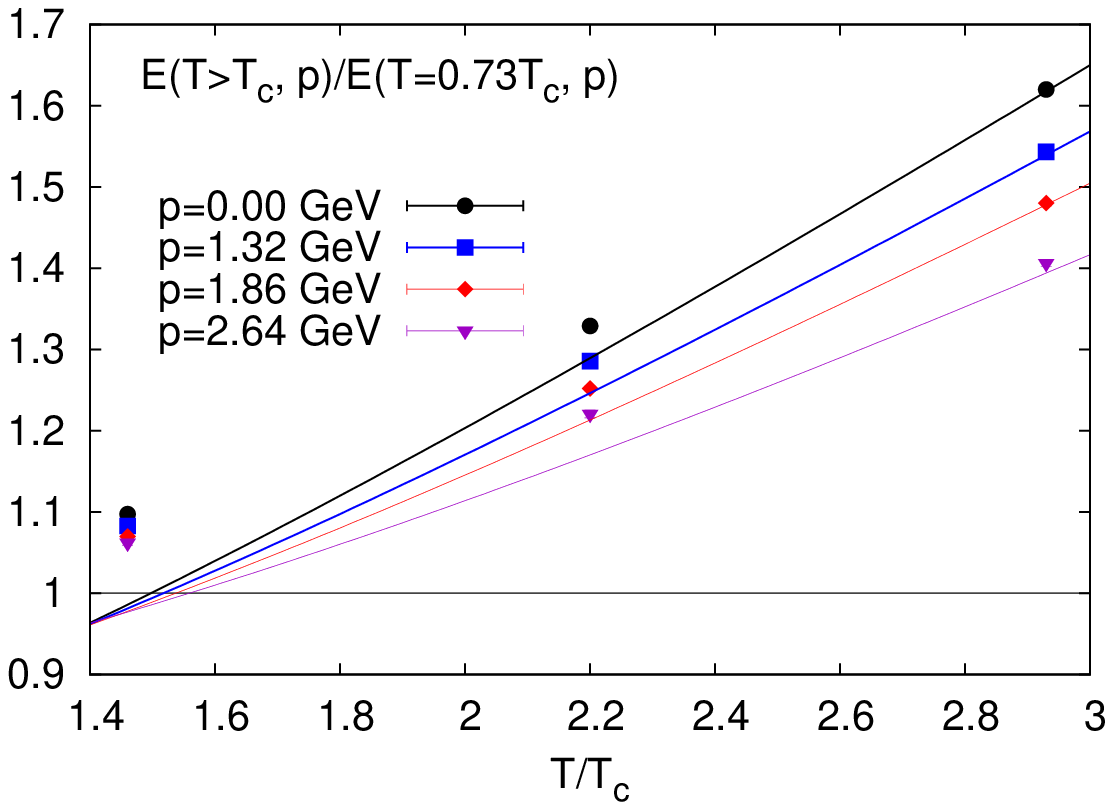} 
    \caption{Left: the dispersion relation of the screening mass in the pseudo scalar channel. The solid lines are linear fits to the data. Right: ratio of $E_{sc}$ at $T>T_c$ to those 
  at $T=0.73T_c$ with same momenta. The solid lines are results from free field theory with charm quark mass $m_c=1.1~$GeV.}
    \label{fig:dispersion_relation_PS}
  \end{center}
  \end{figure}
  
  We show the dispersion relation of the screening mass in the pseudo scalar channel in the left plot of Fig.~\ref{fig:dispersion_relation_PS}. The results are obtained from calculations performed on $128^3\times N_{\tau}$ lattices at 0.73, 1.46, 2.20 and 2.93 $T_c$. The solid  lines denote the dispersion relation obtained by fitting with a linear ansatz. We found
  that the slope of the dispersion relation almost does not change with temperature and only the intercept, i.e. screening mass at rest, 
  changes with $T$. In the right plot of Fig.~\ref{fig:dispersion_relation_PS} we show the ratio of $E_{sc}$ at $T>T_c$ to those 
  at $T=0.73T_c$ with same momenta. The solid lines denote the results from free field theory using a  charm quark mass of $1.1~$GeV~\cite{Ding:2012sp}. At vanishing momentum, 
  the deviations from unity are around 10\% at the lowest temperature and more at higher temperature. This magnitude of deviation is comparable with those obtained from full lattice 
  QCD simulations~\cite{Karsch:2012na}. The ratio becomes closer to unity with increasing momentum
  as in the infinite momentum limit the ratio approaches unity. It is also intuitively expected that the ratio gets closer to the value obtained from free field theory at higher temperatures.

\begin{figure}[htb!]
\centering
\includegraphics[width=0.24\textwidth]{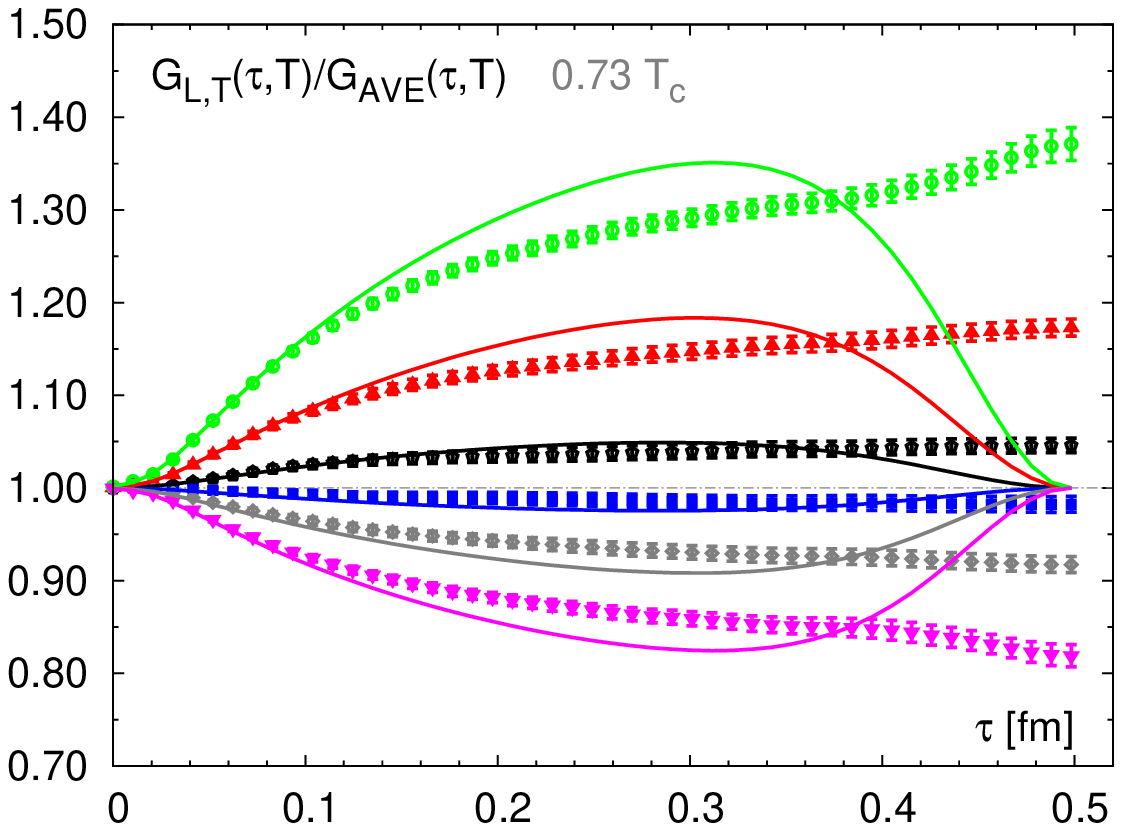}~\includegraphics[width=0.24\textwidth]{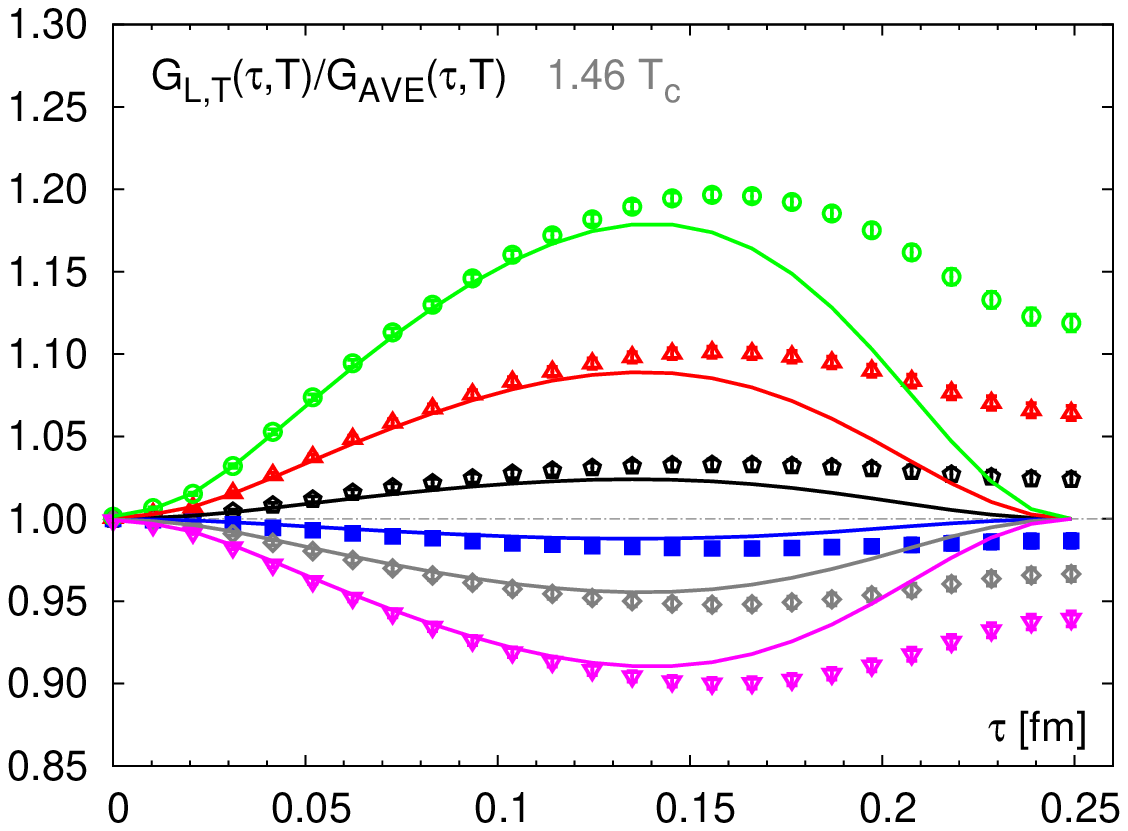}~\includegraphics[width=0.24\textwidth]{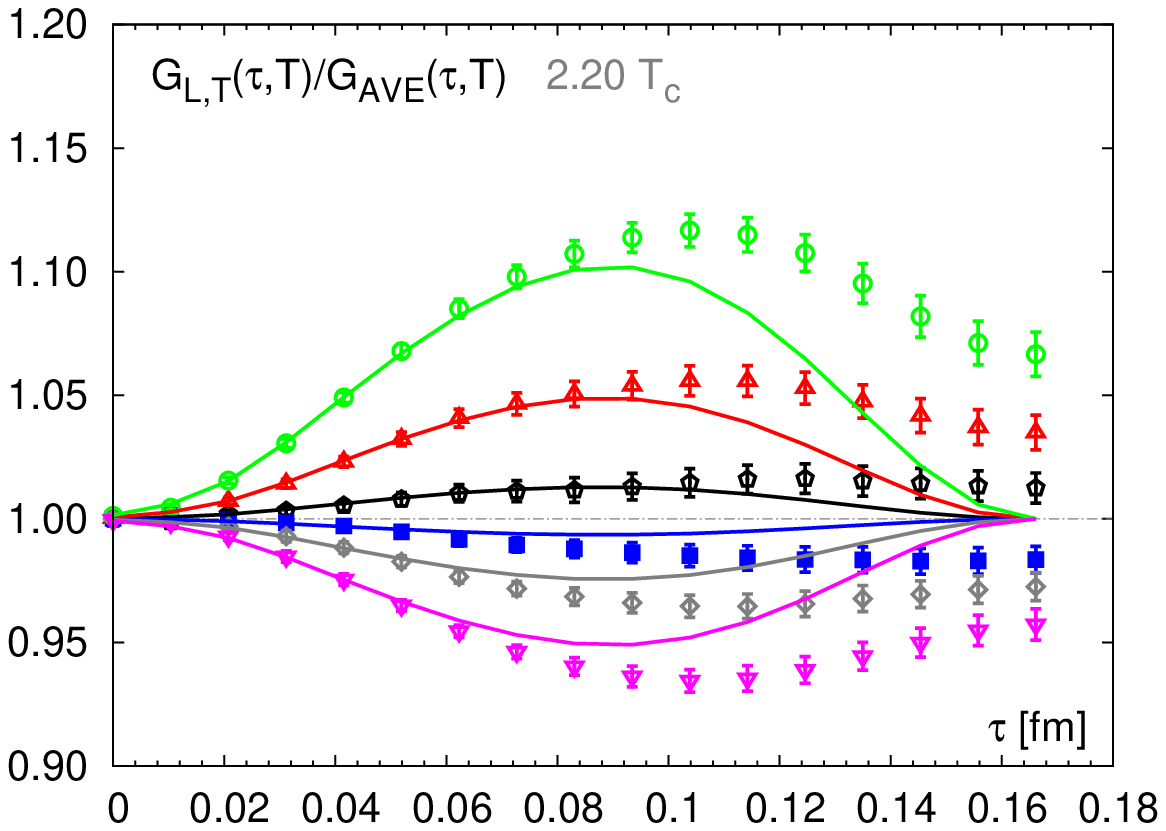}~\includegraphics[width=0.24\textwidth]{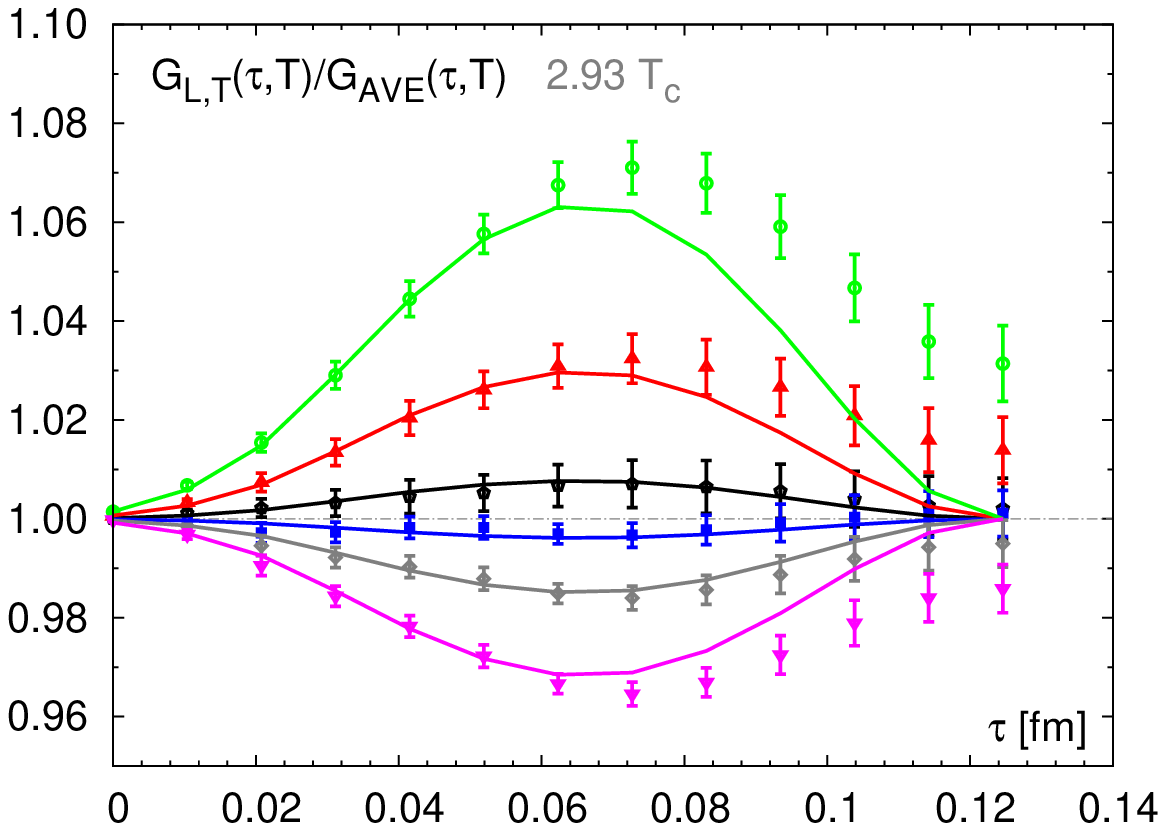}
\caption{Ratios of longitudinal $G_L(\tau,T,p)$ and transverse $G_T(\tau,T,p)$ components of the vector  correlation functions $G_{\rm V_{ii}}(\tau,T,p)$ to 
to the averaged correlator $G_{AVE}(\tau,T,p)$. at 0.73, 1.46, 2.20 and 2.93 $T_c$.  $G_{\rm V_{ii}}(\tau,T,p)$ are obtained using $J_H=\bar{q}\gamma_{i}q$ and $G_{AVE}(\tau,T,p)=G_{\rm V_{ii}}(\tau,T,\vecp)/3= (G_L(\tau,T,\vecp)+ 2G_{T}(\tau,T,\vecp))/3$.
Longitudinal or transverse components are defined such that the momentum are parallel or perpendicular to the $\gamma_i$, e.g. with p=(0.93,0,0) GeV, the longitudinal component is $G_{V_1}$ and the transverse components
are $G_{V_2}$ and $G_{V_3}$. In each plot the six sets of data points from top to bottom correspond to $G_{L}/G_{AVE}$
at p=0.93, 1.86 and 2.97 GeV and $G_{T}/G_{AVE}$ at p=2.97, 1.86 and 0.93 GeV. The solid lines correspond to results obtained from free field theory with charm quark mass $m_c=1.1~$GeV. }
\label{fig:polarized_corr}
\end{figure}

In Fig.~\ref{fig:polarized_corr} we show the ratios of the transverse and longitudinal components to the full vector
correlator. The longitudinal components are always larger than the transverse components at a certain $p$ and $T$. The ratios can be reproduced
by the free field theory with $m_c=1.1~$GeV at short distances~\cite{Aarts:2005hg}.

\begin{figure}[htbl]
\centering
\includegraphics[width=0.32\textwidth]{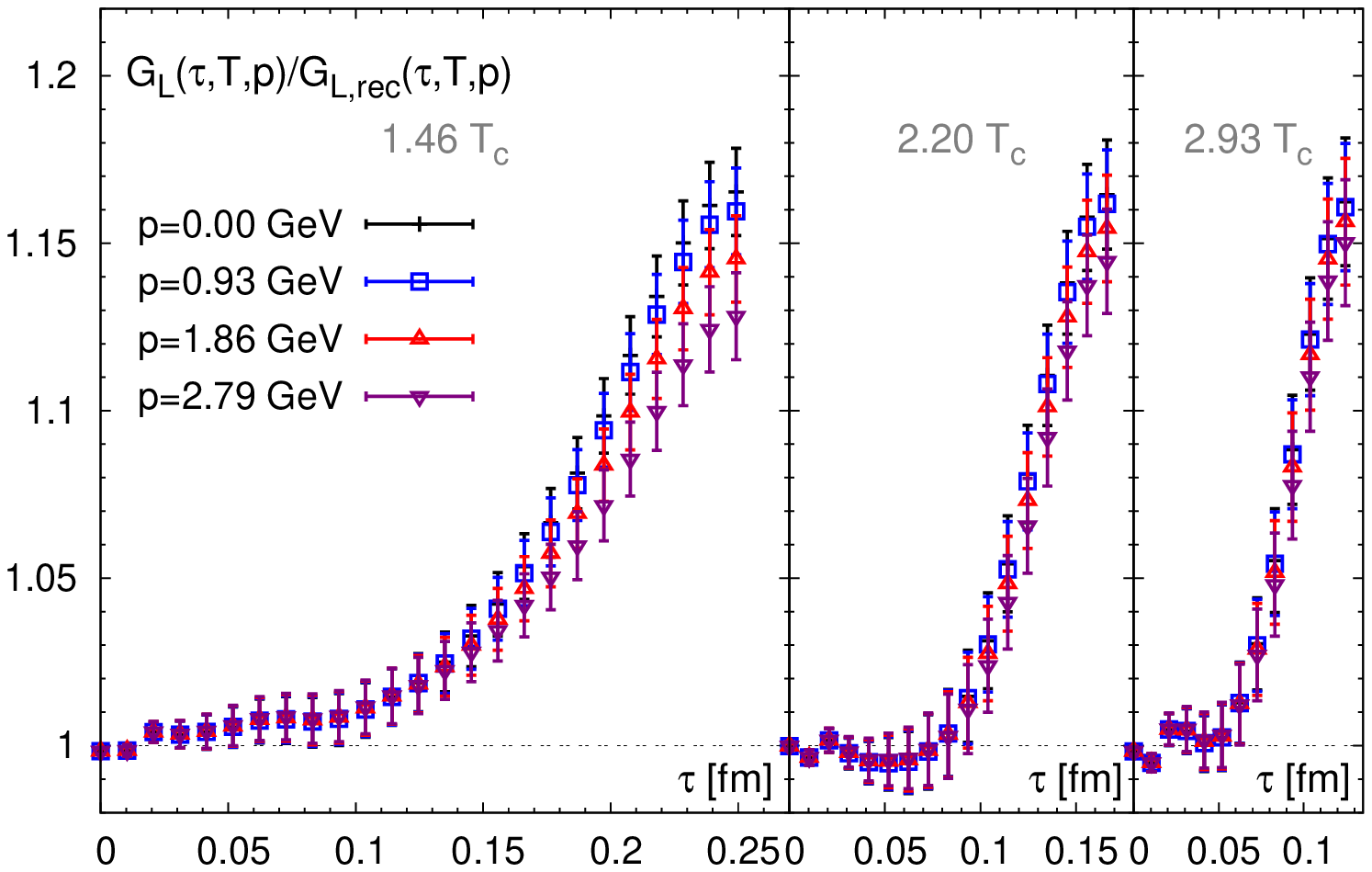}~~\includegraphics[width=0.32\textwidth]{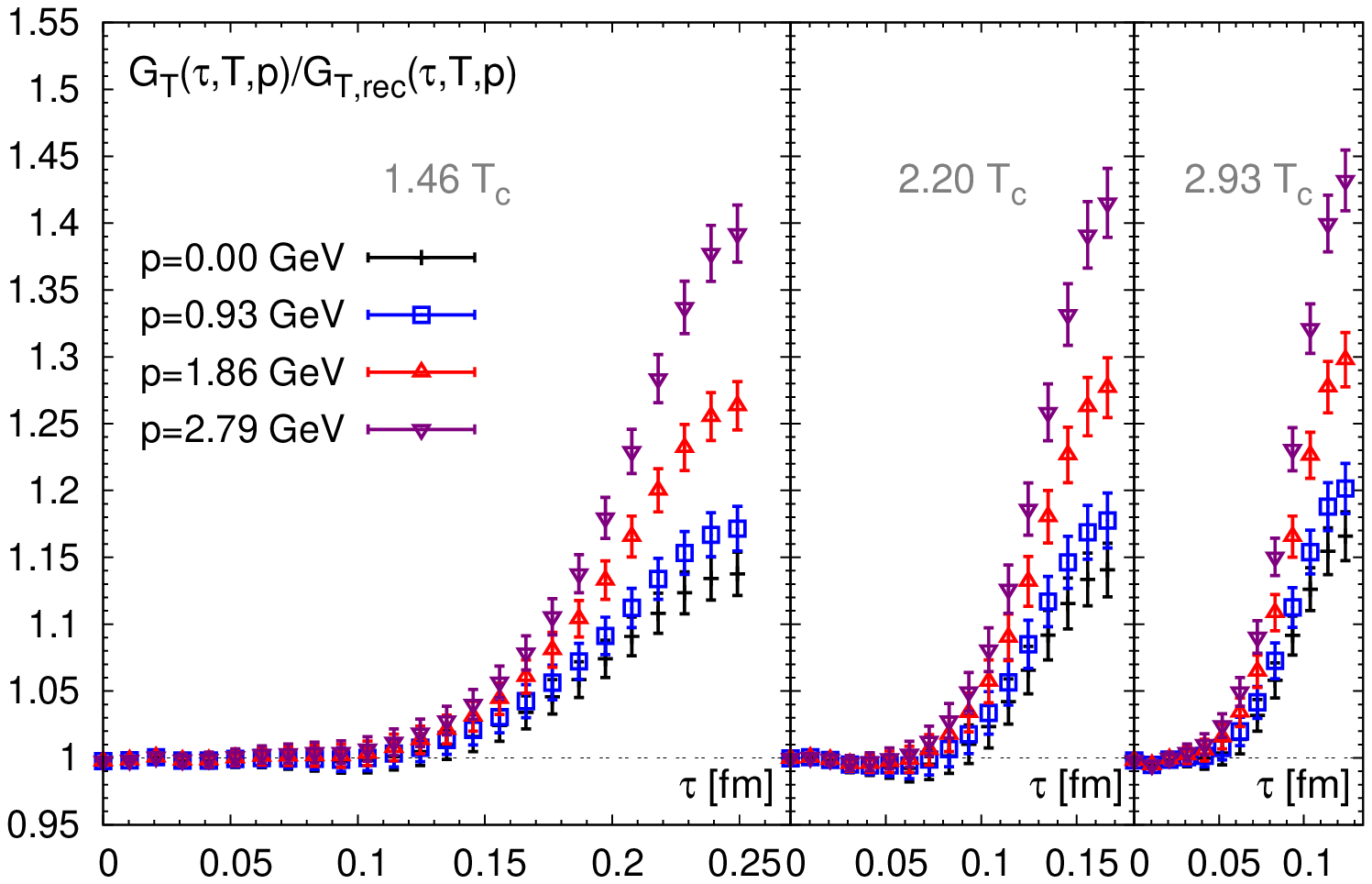}~~ 
\includegraphics[width=0.32\textwidth]{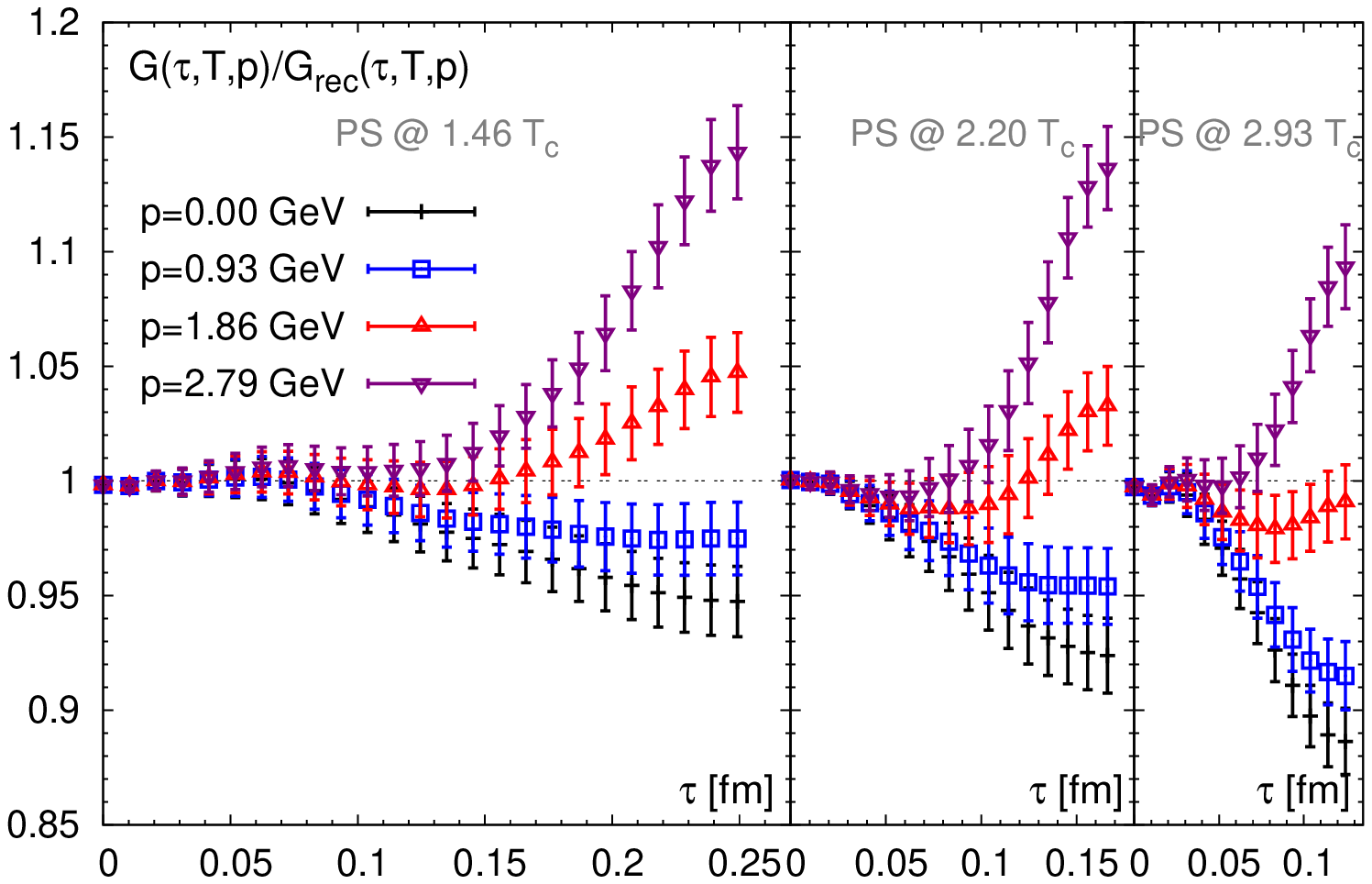} \\
\caption{Ratio of the measured correlator to the reconstructed correlator in the longitudinal (left) and transverse (middle) components in the vector channel and in the pseudo scalar channel (right) at finite momenta.}
\label{fig:GoverGrec}
\end{figure}

In Fig.~\ref{fig:GoverGrec} we show the ratio of measured correlation function to the reconstructed correlation function, which is defined
as 
\be
\frac{G(\tau,T,p)}{G_{rec}(\tau,T,p)}= \frac{\int \mathrm{d}\omega\,\frac{\cosh(\omega(\tau-1/2T))}{\sinh(\omega/2/T)}\, \rho(\omega,T,p)} { \int \mathrm{d}\omega\,\frac{\cosh(\omega(\tau-1/2T))}{\sinh(\omega/2/T)}\, \rho(\omega,T=0.73T_c,p)}.
\ee

In practice one does not necessarily need to evaluate $\rho(\omega,T=0.73T_c,p)$ to get $G_{\rm rec}(\tau,T,p)$ and rather can use Eq.(27) in Ref.~\cite{Ding:2012sp} to 
obtain  $G_{\rm rec}(\tau,T,p)$  directly from the measured correlator $G(\tau,T=0.73T_c,p)$.
One interesting thing to see is that the ratio $G_{\rm L}/G_{\rm L,rec}$ almost does not change with momentum at each temperature 
while in the transverse components of the vector channel and in the pseudo scalar channel the momentum effects bring
ratios up.
This may be an indication
of different non-trivial small frequency behavior as suggested in Ref.~\cite{Hong:2010at}.  Note that calculations in 
Ref.~\cite{Hong:2010at} are  done for the light quark modes thus it would be more suitable to have a follow
up study on the momentum dependences of the light correlation functions after Ref.~\cite{Ding:2010ga}.

\begin{figure}[htbl]
\centering
\includegraphics[width=0.33\textwidth]{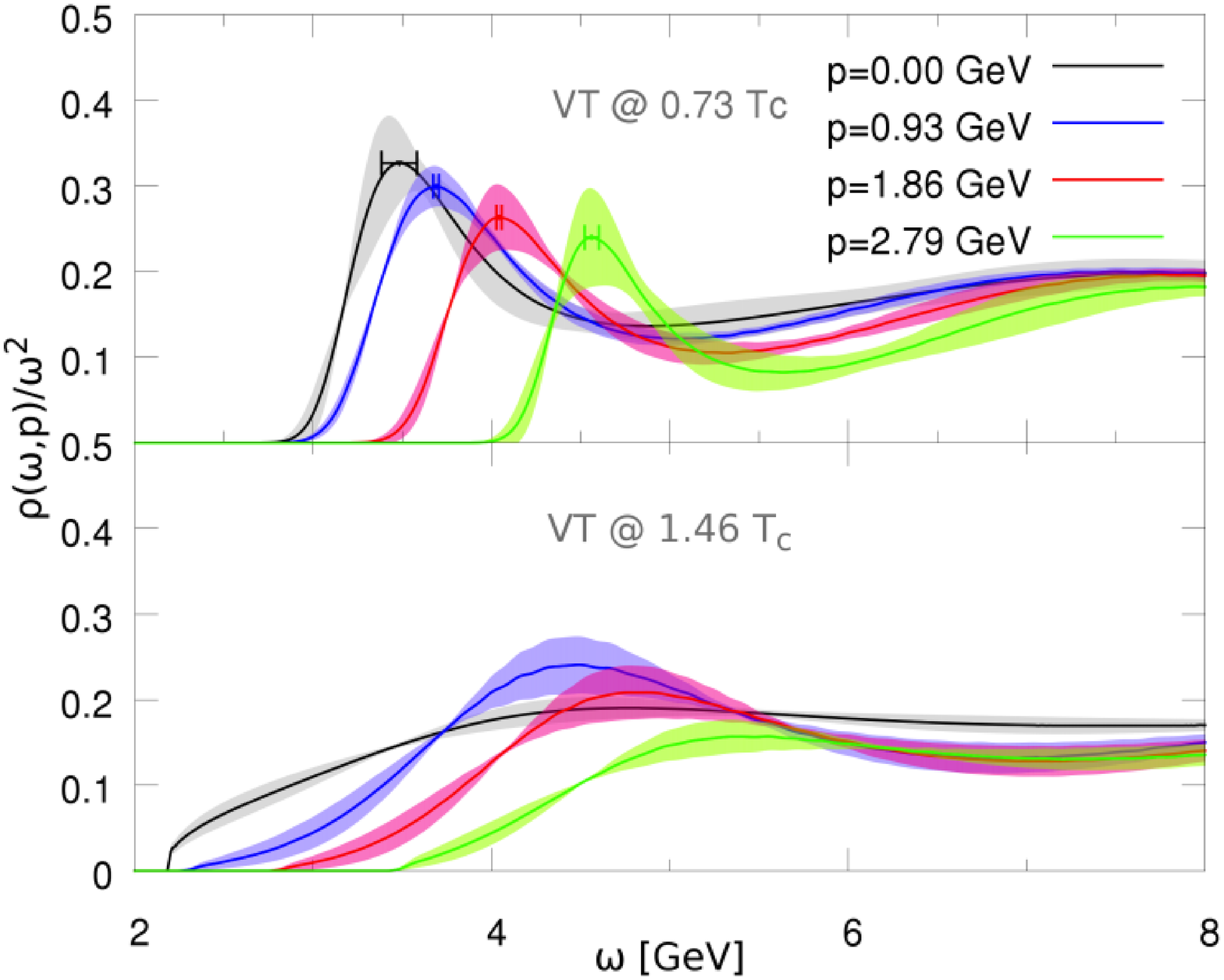}~~~~~~\includegraphics[width=0.33\textwidth]{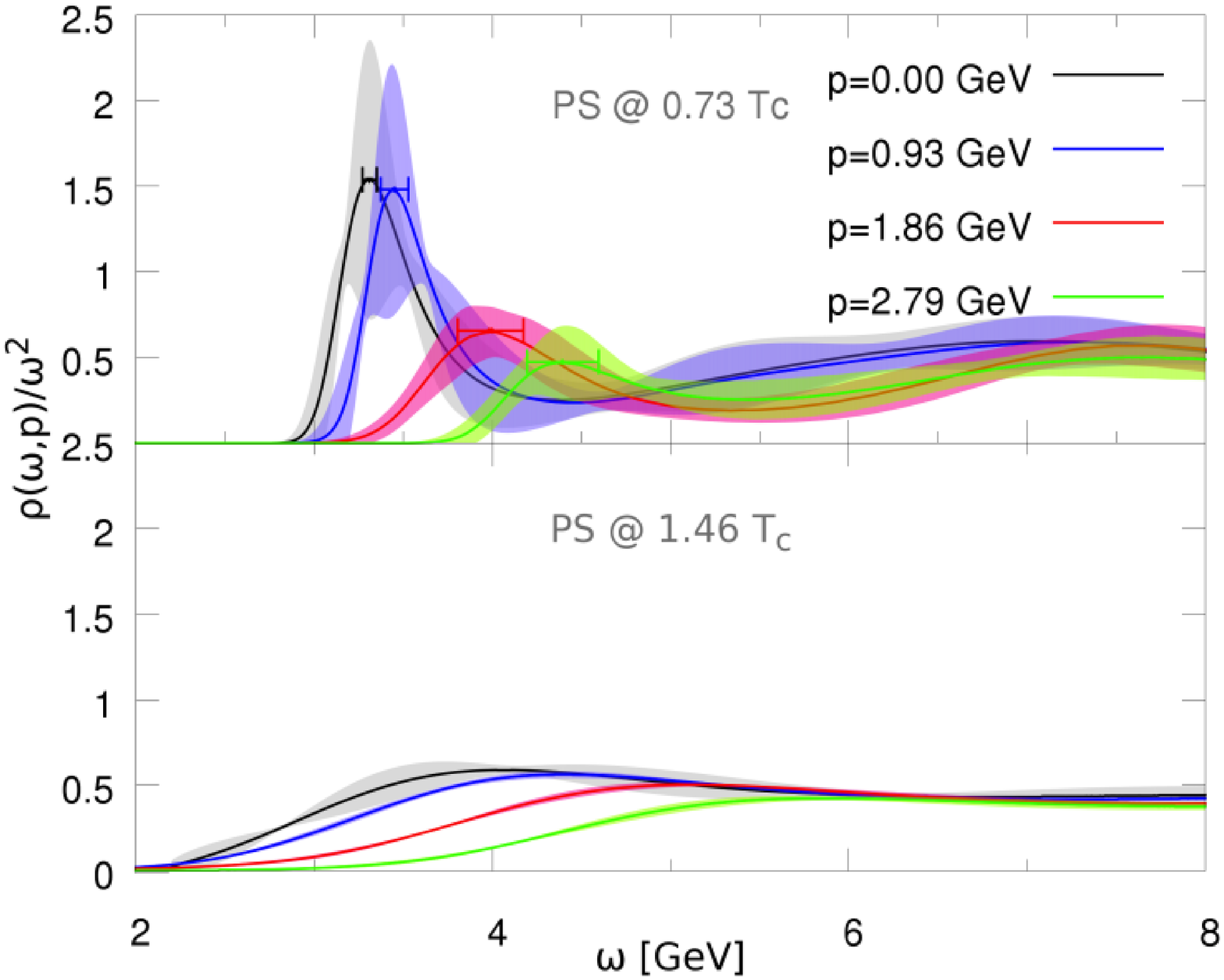} 
\caption{Momentum dependences of spectral functions in the vector (left) and pseudo scalar (right) channels at $T=0.73T_c$ and 1.46$~T_c$.}
\label{fig:spf}
\end{figure}

In Fig.~\ref{fig:spf} the spectral functions extracted using the Maximum Entropy Method~\cite{Asakawa:2000tr} 
in the transverse components of the vector channel and in the pseudo scalar channel are shown.
The bands correspond to the statistical uncertainties and the horizontal error bars 
give the uncertainties of the peak locations at 0.73 $T_c$. The change of the peak locations at $T<T_c$ in both channels are well aligned with the study
on the dispersion relation of screening masses.  At 1.46~$T_c$ spectral functions in both channels get attenuated in the small $\omega$ region and the threshold
moves to the large $\omega$ region.
\section{Summary}

We have studied the momentum dependences of charmonium properties by investigating on the
temporal and spatial correlation functions. As the static $\eta_c$ and $\Jpsi$ are found to dissolve already at  $T\gtrsim 1.5~T_c$~\cite{Ding:2012sp},
to study the properties of charmonium moving respect to the heat thermal bath, Monte Carlo simulations 
at temperature values below 1.5 $T_c$ are crucially needed to examine the sequential dissociation
picture of charmonia system as well as the possible modifications of charmonium bound states at finite momentum~\cite{futhure}.


\end{document}